# Self-protecting aqueous lithium-ion batteries with smart thermoresponsive separators

Yuewang Yang[1,3], Zhaowen Bai[2,3], Sijing Liu[1,3], Yinggang Zhu[2], Jiongzhi Zheng[1], Guohua Chen[2] and Baoling Huang[1]

[1]Department of Mechanical and Aerospace Engineering, The Hong Kong University of Science and Technology, Clear Water Bay, Kowloon, Hong Kong SAR, China.

[2]Department of Mechanical Engineering, The Hong Kong Polytechnic University, Hung Hom, Kowloon, Hong Kong, China

[3]These authors contributed equally to this work.

Corresponding author. Email: guohua.chen@polyu.edu.hk  mebhuang@ust.hk

Capacity degradation and destructive hazards are two core challenges for lithium-ion batteries at high temperatures, which need to be solved urgently. Adding flame retardants or fire extinguishing agents can only achieve one-time self-protection in case of emergency overheating. Herein, smart self-protecting aqueous lithium-ion batteries were developed using thermoresponsive separators through in-situ polymerization on the hydrophilic separator. The thermoresponsive separator will close the lithium ions transport channel at high temperatures and reopen when the battery cools down; more importantly, the transition is reversible. We studied the lithium salts influence on the thermoresponsive properties of the hydrogels and selected suitable lithium salt ($LiNO_3$) and concentration (1 M) in the electrolyte to achieve self-protection without sacrificing battery performance. In addition, the shut-off temperature can be tuned by adjusting the hydrophilic and hydrophobic moiety ratio in the hydrogel according to actual demands. This self-protecting lithium-ion battery shows promise for smart energy storage devices with safe and extended lifespan.

Driven by the rapidly growing mobile energy storage demands, such as electric vehicles and portable electronic devices, the development of lithium-ion batteries with higher energy density, improved safety and longer cycle life have received continuous great attention in recent years[1–3]. Traditional non-aqueous lithium-ion batteries using toxic flammable organic electrolytes have been facing the problem of pollution and unsafety[4]. Aqueous rechargeable lithium-ion batteries (ARLB) with intrinsically safer, cheaper and environment-friendly aqueous electrolytes of higher conductivities are thus promising alternatives for sustainable next-generation high-safety battery systems[5]. A number of ARLB systems with different electrode materials have been explored since the first ARLB proposed by Dahn and co-workers[6] in 1994. Among various cathode materials, $LiMn_2O_4$ (LMO) has been considered as the most promising cathode material due to its excellent chemical stability and electrochemical cycling stability in aqueous electrolytes[5,7,8]. However, there are only a few suitable candidates for the anode materials, as the operation potential is limited at 2-3 V versus $Li^+/Li$[9] due to the narrow electrochemically stable window of aqueous electrolytes. These anode candidates can be mainly divided into two categories: vanadium-based materials[10] (e.g., $VO_2$, $LiV_3O_8$, $Li_xV_2O_5$ and $H_2V_3O_8$) and polyanionic-based materials[11] (e.g., $LiTi_2(PO_4)_3$ and $TiP_2O_7$). Unfortunately, the poor cycling performance due to the degradation of crystal structures and vanadium dissolution in aqueous electrolytes has hindered the development of vanadium-based anodes[9]. The NASICON-type $LiTi_2(PO_4)_3$ (LTP) exhibits relatively high specific capacity, proper operating potential (2.5 V versus $Li^+/Li$, -0.5 V versus hydrogen electrode (NHE)), and high stability in aqueous electrolytes, and thus is considered as a promising anode material[11]. However, it needs to be coated with conductive carbon (C-LTP) in applications due to its low electrical conductivity[7,8].

Despite many merits of ARLB, the fast capacity degradation at high temperatures caused by reduction of active materials and the accelerated side reactions of electrolytes is still a serious problem, which heavily reduces battery life[12–15]. Self-ignition or even explosion may happen if heat continues to accumulate[16]. Some traditional lithium-ion batteries solve these safety problems through adding flame retardants into the electrolyte, which is accompanied by deterioration in cycling performance, or one-time protection such as extinguishing agents, shutdown current collectors and fused disconnect switches[17–19]. Therefore, smart and reversible self-protection strategies need to be further developed to design aqueous lithium-ion batteries with better high temperature adaptability and safety guarantee.

The poly(N-isopropylacrylamide) (PNIPAm)-based hydrogel is a smart thermoresponsive material, in which the polymer chains can change from hydrophilic to hydrophobic state above a certain temperature and release the swelling solution while the phenomena will reverse if the temperature is back below that lower critical solution temperature (LCST)[20,21]. PNIPAm-based hydrogels have been widely used in drug delivery systems[22], switching memory devices[23], self-healing materials[24], and so on. Recently, some researchers applied this smart hydrogel to energy storage devices, including supercapacitors[25–27] and Zinc-ion batteries[28,29], which may endow the devices with self-protection capability at high temperatures.

However, hydrogels used in batteries need to meet higher requirements due to the complex electrochemical reactions and ion migration at the electrolyte-electrode interfaces, leading to limitations on the battery performance. For example, reported thermoresponsive Zinc-ion batteries based on smart hydrogels could only adopt low-concentration electrolytes (e.g., 0.3 M $ZnSO_4$[28,29]) due to the destructive effects of the kosmotropes salts on the hydrogel properties including the thermal responsivity, which limits the cycling performance[30–33] of the battery. In contrast, chaotropes ions, such as $NO_3^-$, can maintain the thermoresponsive properties of the hydrogels even at high concentrations[34–36]. However, $NO_3^-$ is instable in the Zn deposition/dissolution processes[37,38] and therefore zinc batteries with $Zn(NO_3)_2$ exhibit very low Coulombic efficiency. Interestingly, in aqueous lithium-ion batteries, the inclusion of $NO_3^-$ does not deteriorate the battery performance. $LiNO_3$ can work as well as other lithium salts, such as $Li_2SO_4$ and LiTFSI, in lithium ion batteries[5,39–41].

Here, we developed a smart self-protecting aqueous lithium-ion battery using thermoresponsive separators and 1 M $LiNO_3$ electrolytes. The influence of different salts on the thermoresponsive properties of the hydrogel has been investigated. The salts will decrease the LCST of the hydrogels, and the higher salt concentrations, the lower LCST. More importantly, the species of salts have a significant impact on the decrease of the LCST. The chaotropes salt, $LiNO_3$, was chosen as the lithium salt in our ARLB due to its minor effect on thermoresponsive properties of the hydrogels, and appropriate concentration (1 M) was used to maintain good battery performance. In addition, the shut-off temperature of the battery can be tuned through adjusting the ratio of more hydrophilic groups (acrylamide, Am) to N-isopropylacrylamide (NIPAm) in the hydrogel. The thermoresponsive separator was evaluated through Impedance tests, SEM, Raman, DSC, and water contact angle tests. Finally, $LiMn_2O_4$/carbon coated $LiTi_2(PO_4)_3$ (LMO/C-LTP) pouch-cell batteries using thermoresponsive separators were demonstrated, and it exhibited almost the same specific capacity (110 mAh/g at 1C) and cyclic stability as the ARLB with hydrophilic separators. The reversible self-protecting property of the battery was tested in several consecutive heating-cooling cycles and in a visual LED lighting test.

The self-protection mechanism is shown in Fig. 1. At room temperature, the hydrogel on the surface of the hydrophilic separator is swelling (open channel) and hydrophilic, which allows the ions to migrate freely. When the temperature is above the LCST, the hydrogel turns to the hydrophobic state and squeezes the electrolyte out (closed channel), which impedes the transport of ions in the devices and prohibits heat accumulation. This process is completely reversible, and the devices shut down at high temperature can recover its normal operation when they cool down below the LCST.

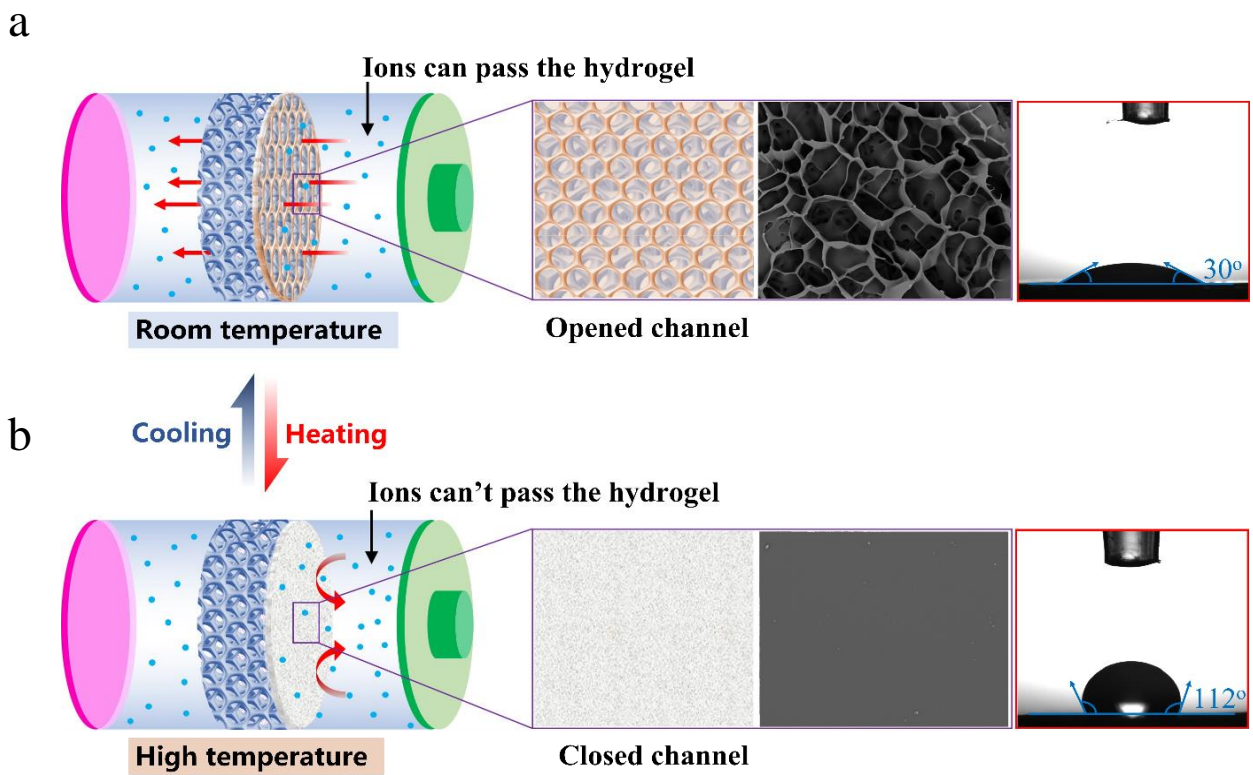

Fig. 1. (a) Opened channel and hydrophilicity of the separator below LCST. (b) Closed channel and hydrophobicity of the separator above LCST.

An in-situ free radical polymerization approach was employed to prepare the thermoresponsive separator with N-isopropylacrylamide (NIPAm) and acrylamide (Am) as monomers, ammonium persulfate (APS) as the initiator, tetramethylethylenediamine (TEMED) as the catalyst and methylene-bis-acrylamide (MBAA) as the cross-linker. The hydrogel grew on the hydrophilic separator and formed a thin and uniform thermoresponsive separator with hydrophilic separator as the matrix and hydrogel as the smart switch (Fig. 2a, Fig. S3). The synthesis of the hydrogels is shown in Fig. 2b and the final composition of the hydrogels is controlled by the feed ratio. The hydrogels are composed of a certain amount of NIPAm and a series weight percentages of Am, ranging from 0 to 30wt.%, which are denoted as $PNIPAm_{100}$ to $P(NIPAm_{70}$-co-$Am_{30})$, respectively. Owing to the low surface tension of the aqueous solution on the hydrophilic separator, the coated hydrogel on the hydrophilic separator is very uniform and the thickness is about 60 μm (Fig. S3).

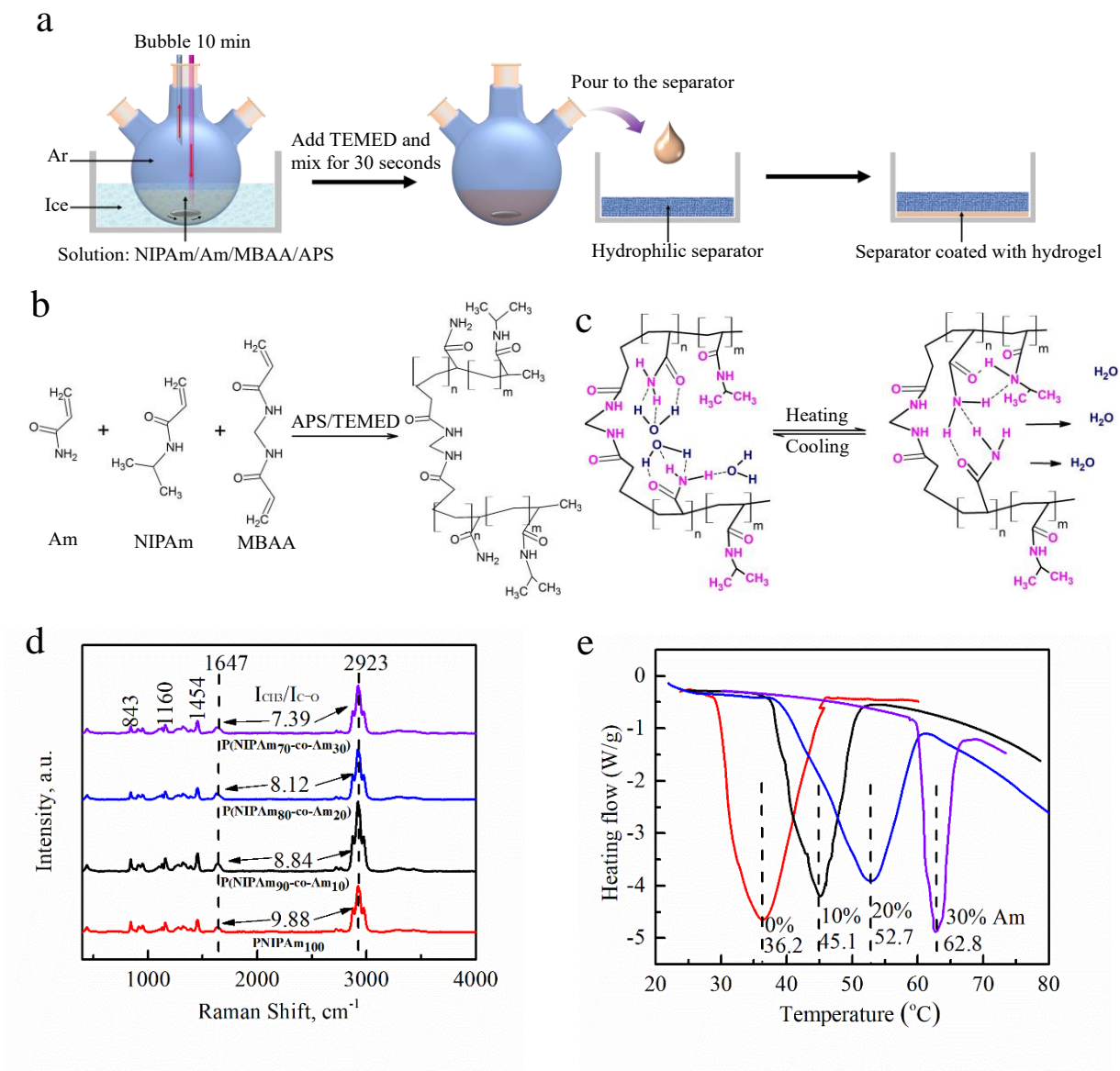

Fig. 2. (a) Preparation process flow of the thermoresponsive separator. (b) Structure formula. (c) Mechanism of the reversible opened channel-closed channel transition of hydrogel in the heating-cooling conversion process. (d) Raman spectra of freeze-dried hydrogels with different Am ratios. (e) LCST of hydrogel with different Am ratios in water.

The mechanism of the thermoresponsive behaviors of the hydrogel is illustrated in Fig. 2c. The repeating units in the polymer have both hydrophobic and hydrophilic moieties, and the hydrophilic moieties will form hydrogen bonds with water molecules below the LCST, which corresponds to the swelling of the network for the hydrogel. In contrast, the hydrogen bonds between water molecules and hydrophilic moieties are broken above the LCST, and the hydrophobic interactions among the hydrophobic backbone in the polymer become dominant, which corresponds to the process of deswelling and squeezing out water molecules.

Above the LCST, molecular chains in the polymer, for example, pure PNIPAm, curl into clusters and precipitate out of solution irregularly. Based on this feature, we added MBAA as the cross-linker agent to convert the chain polymer to the covalent chemical cross-linked three-dimensional network. The covalent chemical cross-linked hydrogels will squeeze out the aqueous electrolyte and keep roughly the original structure, rather than disorder phase separation as in chain polymers above the LCST, which benefits designing the structure that keeps its shape (a completely closed membrane) under high temperature (Fig. S4).

The LCST can be tuned by adjusting the ratio of hydrophilic and hydrophobic groups in the hydrogel. The Am monomers with high contents of C=O and N-H shows higher hydrophilicity compared with NIPAm, and the hydrogel with high proportion of Am can form more hydrogen bonds with water. With the proportion of Am rising from 0% to 30wt%, higher energy is required to break the increased hydrogen bonds, and thus the LCST increases from 34.6 °C to 62.8 °C (Fig. 2d). Here, we used the tip of the DSC peak to represent the phase separation temperature.

The hydrogels of different compositions were freeze-dried and characterized by Raman and Fourier transform infrared (FTIR) spectra. The Raman results are showed in Fig. 2e. The peak at 1647 cm$^{-1}$ can be assigned to C=O stretching present in both monomers of NIPAm and Am, and the peak at 2923 cm$^{-1}$ is correlated with a group present only in monomers of NIPAm[42]. The $I_{CH3}/I_{C=O}$ ratios of PNIPAm$_{100}$, P(NIPAm$_{90}$-Am$_{10}$), P(NIPAm$_{80}$-Am$_{20}$) and P(NIPAm$_{70}$-Am$_{30}$) are 9.88, 8.84, 8.12 and 7.39, respectively. The decrease in the ratios indicates an increase in the portion of hydrophilic monomers Am in the hydrogel. The peaks at 1408 and 1656 cm$^{-1}$ corresponding to the C=C and –C=C stretching vibrations in NIPAm and Am monomers could be clearly seen in the FTIR spectra (Fig. S5). These peaks disappear after the polymerization, which confirms the transition of monomers into polymers[43].

Since the hydrogel is used in electrolytes of the battery, which contains conductive ions, the influence of ions on the thermoresponsive property must be considered. The change of LCST with increasing salt concentrations is shown in Fig. 3a. The LCST of hydrogels at different LiNO$_3$ concentrations of 0, 1, 2, 3, and 4 M are 52.7, 47.5, 41.3, 38.1, and 30.9 °C, respectively. When the salts are added to the hydrogel, the water molecules will be polarized by the ion species and the interactions between H in the water and the O, N in the polymer are weakened, which ultimately leads to decrease of the LCST[36]. In addition, the enthalpy of phase separation of the hydrogel in water (506.9 J/g) is larger than that in LiNO$_3$ electrolytes (below 150 J/g) due to the weaker hydrogen bonds in the electrolytes (ions in water) (Fig. 3a). The degree of LCST reduction is related to the Hofmeister series and salt concentration[34,36,44–46].

$$SO_4^{2-} > PO_4H^{2-} > F^- > CH_3COO^- > Cl^- > Br^- > I^- > NO_3^- > ClO_4^- > SCN^-$$

<div align="center">Hofmeister series</div>

The kosmotropes ions (left in the Hofmeister series) will drastically reduce the LCST. In contrast, the chaotropes ions (right in the Hofmeister series) have much weaker effects on the hydrogel. In addition, anions appear to have much larger effects than cations[47] and we mainly discuss the impact from anions here.

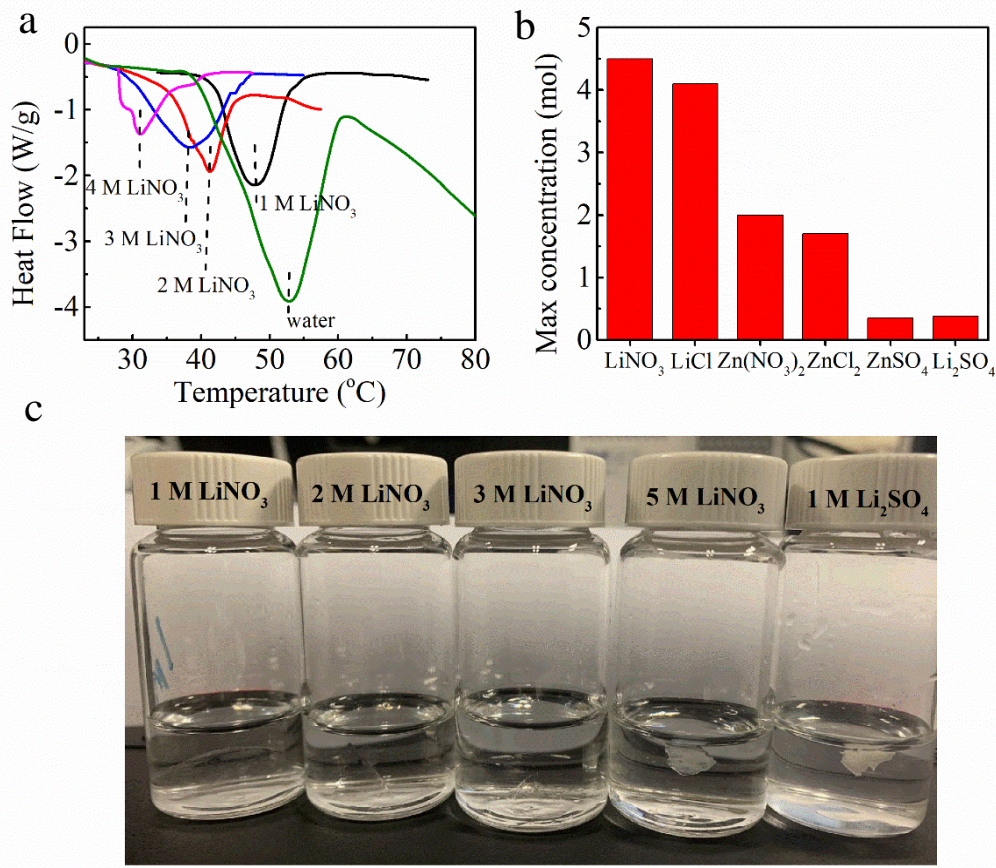

Fig. 3. (a) LCST of P(NIPAm$_{20}$-Am$_{80}$) in LiNO$_3$ electrolytes of different concentrations, and the area enclosed by the peak represents the absolute value of the enthalpy. (b) The maximum concentrations of the hydrogel can maintain the swelling state at room temperature for different salts. (c) Picture of swelling hydrogel at low concentration of LiNO$_3$ solution and deswelling hydrogel at high concentration of LiNO$_3$ solution and 1M Li$_2$SO$_4$ at the room temperature.

The maximum concentrations of different salts at which the hydrogel can maintain swelling at room temperature are listed in Fig. 3b. The SO$_4^{2-}$ based electrolyte would shrink the hydrogel at room temperature when the concentration is higher than 0.3 M, whereas the NO$_3^-$ based electrolyte allows the hydrogel to remain swelling even at the concentration of 3 M (Fig. 3c).

The thermoresponsive separator exhibits hydrophilicity at room temperature and displays a small water contact angle (WCA) of about 30 degrees. In contrast, the hydrophobic interactions

among –O=C-NH-R replace the hydrophilic interactions between –O=C-NH-R and water molecules as the dominant part above the LCST (Fig. 2c), and the conversion from hydrophilic to hydrophobic state is manifested by the increase of WCA from 30 ° to 100 ° (Fig. 4a). As shown in Fig. 4b, the contact angle increases with the increasing temperature and the sudden increase at around 50 °C corresponds to the LCST of PNIPAm$_{80}$-Am$_{20}$ in the 1 M LiNO$_3$ electrolyte. Besides the change of wettability, the structure of the hydrogel also undergoes a tremendous variation from the swelling state to the deswelling state. The scanning electron microscopy (SEM) images of the hydrogels at the swelling state and deswelling state are shown in Fig. 4d, e and f. The swelling hydrogel has many opened channels that allow for the transport of ions, while these channels are closed at the deswelling state, which cuts off the transport of ions between the cathode and anode. The dual effects of repelling to aqueous electrolytes caused by hydrophobicity and closed channel in the polymer completely disable the ions transport and the corresponding electrochemical reactions. This phenomenon is reversed when the polymer is cooled down.

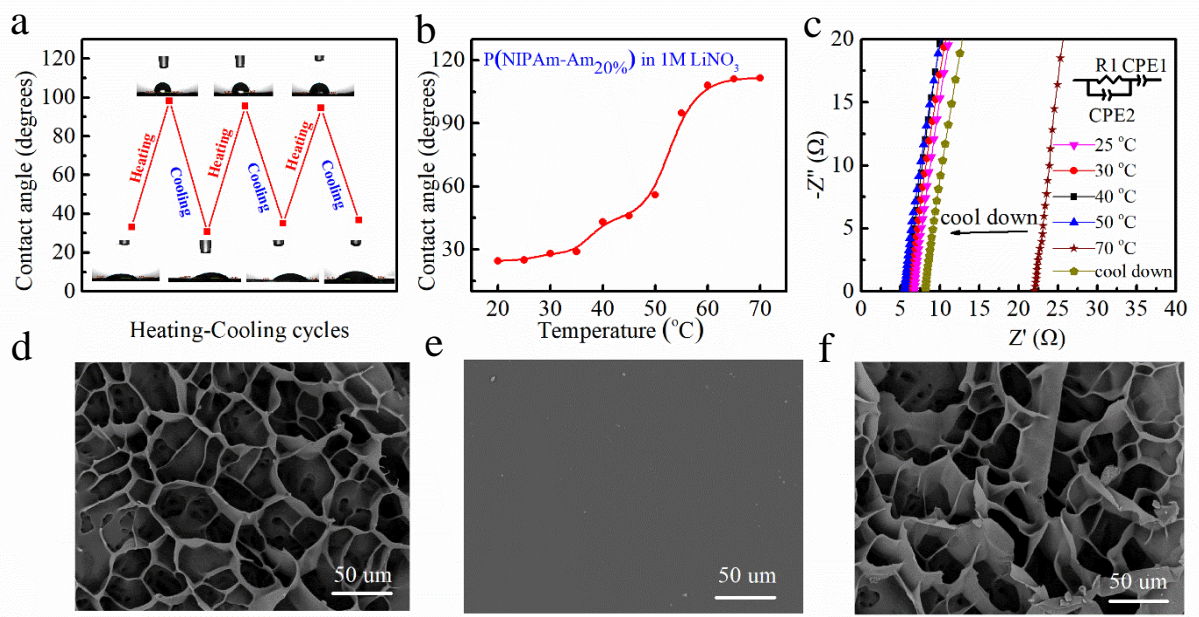

Fig. 4. (a) Variations of the contact angles of the thermoresponsive separator at different temperatures in heating-cooling cycles. (b) The contact angel of P(NIPAm$_{80}$-Am$_{20}$) in 1 M LiNO$_3$ at different temperatures. (c) EIS results of the thermoresponsive separator at different temperatures. (d) SEM image of hydrogel in swelling state and (e) deswelling state at high temperature. (f) SEM image of the hydrogel after cooling down.

The ability to hinder ion transport at the deswelling state was characterized through the impedance measurement (Fig. 4c). For our thermoresposive separator, the impedance is similar to the hydrophilic separator when the temperature is lower than 60 °C, indicating that the in-situ grown hydrogel has a low impedance. When the temperature increases, the viscosity of the electrolyte decreases and ion migration becomes faster, leading to decrease in impedance. However, when the temperature is above 60 °C, the impedance increases by over an order of magnitude, which corresponds to the deswelling state (Fig. 4e). After cooling down to the room temperature again, the impedance decreases again to the initial level. In contrast, the impedance of hydrophilic separator gradually decreases with increasing temperature (Fig. S7).

Based on the above-mentioned study on the effects of the ion species on the performance of aqueous lithium-ion batteries and the thermoresponsive behavior of thermoresponsive separator, we designed and assembled pouch-cell batteries of smart aqueous lithium-ion batteries. The batteries adopted 1 M LiNO$_3$ as the electrolyte, thermoresponsive separator as the separator, lithium manganese oxide as the positive electrode, and carbon coated lithium titanium phosphate as the negative electrode. For comparison, we also adopted the normal hydrophilic separator as a reference sample. Galvanostatic Charge-Discharge (GCD) tests were conducted to evaluate the battery performance. The LMO/C-LTP battery based on thermoresponsive separator showed similar charge/discharge plateau and specific capacity compared with the battery with the hydrophilic separator (Fig. 5a). Fig. 5b exhibits the rate performance of batteries with the thermoresponsive separator and hydrophilic separator at current ranging from 0.5C to 4C.

The delivered specific capacity is 110 mAh/g at 0.5C, and it can still reach 80 mAh/mg at 4C. Furthermore, the cyclic performances of the aqueous lithium-ion batteries with two separators were tested at 1C (Fig. 5c). After 100 cycles, the batteries could still maintain the specific capacities of 74.42 mAh/g and 74.64 mAh/g for hydrophilic separator and thermoresponsive separator, respectively. Overall, these results prove that the modified separator would not deteriorate the battery performance due to negligible effect on ion migration from the thin and uniform hydrogel attached on the hydrophilic separator. The ARLB using 1 M $Li_2SO_4$ electrolytes exhibits the same cyclic performances as the ARLB using 1 M $LiNO_3$ electrolytes (Fig. S8)

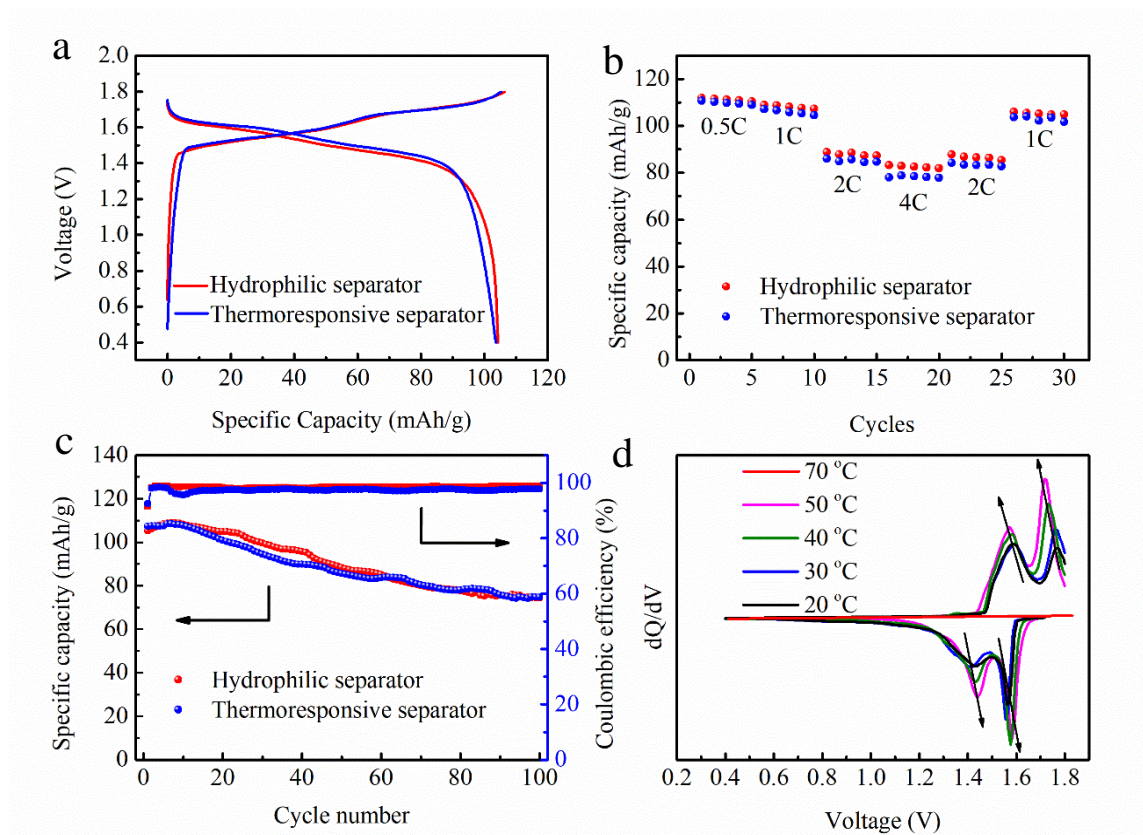

Fig. 5. (a) Galvanostatic charge-discharge (GCD) curves of the two batteries with thermoresponsive and hydrophilic separators. (b) Rate performance of the two batteries with hydrophilic and thermoresponsive separators. (c) Cyclic performance at 1C for two batteries with hydrophilic separator and thermoresponsive separator. (d) Differential specific capacity vs. voltage (dQ/dV) curves of the battery with thermoresponsive separator.

Fig. 5d represents the differential specific capacity dQ/dV curves of the lithium-ion batteries with the thermoresponsive separator in the temperature range from 20 °C to 70 °C. Two reversible redox peaks could be observed for the battery below 50 °C, corresponding to the two-step intercalation/deintercalation of Li+ into/from the different lattice sites in $LiMn_2O_4$[48]. The anodic and cathodic peaks become higher as the temperature increased due to the faster movement of ions in the electrolyte and electrodes at higher temperature. In addition, when the temperature increases from 20 °C to 50 °C, the two anodic peaks move from 1.766 to 1.714 V and from 1.602 to 1.571 V, respectively. Meanwhile, the two cathodic peaks move from 1.556 to 1.676 V and from 1.419 to 1.441 V, respectively. The peak shifts mean that the polarization of the battery becomes smaller as the temperature increases. However, the peaks disappear, and the curve becomes a horizontal line with a value of zero because the battery has no capacity left above the LCST of the hydrogel. For comparison, we also tested the hydrophilic separator based batteries at different temperatures (Fig. S8) and the height of the redox peaks is still increasing at 70 °C.

To further analyze the self-protecting property of the thermoresponsive separator, Electrochemical impedance spectroscopy (EIS) test of full battery was conducted at different temperatures with the frequency range of $10^5$-0.1 Hz. The inset in Fig. 6a is the equivalent circuit model used to fit the EIS data and the fitting results are given in Table S1. The *Rs* is the electrolyte resistance, *Rct* is the charge-transfer resistance at two electrode-electrolyte interfaces, *Zw* is the interfacial diffusion resistance (Warburg impedance) that related to the ion diffusion in the electrode materials, and the *CPE* is the double layer capacitance[49]. Upon heating from 30 °C to 50 °C, the *Rct* of both modified separator and hydrophilic separator based batteries become smaller with the impedances decrease from 119.4 to 69.47 Ω and from 111.7 to 92.6 Ω, respectively (Fig. 6a, 6b). This is because the ion migration rate becomes faster at higher temperatures. However, the impedance trends become opposite for the two separators when the batteries are heated to 70 °C. The resistance of hydrophilic separator based battery maintains a consistent decrease trend at 70 °C, from 92.6 Ω to 66 Ω (Fig. 6b). In contrast, the thermoresponsive separator based battery loses the battery characteristic, where a circle at high frequency and a diagonal line at low frequency becomes a diagonal line. This is a typical characteristic of a capacitor instead.

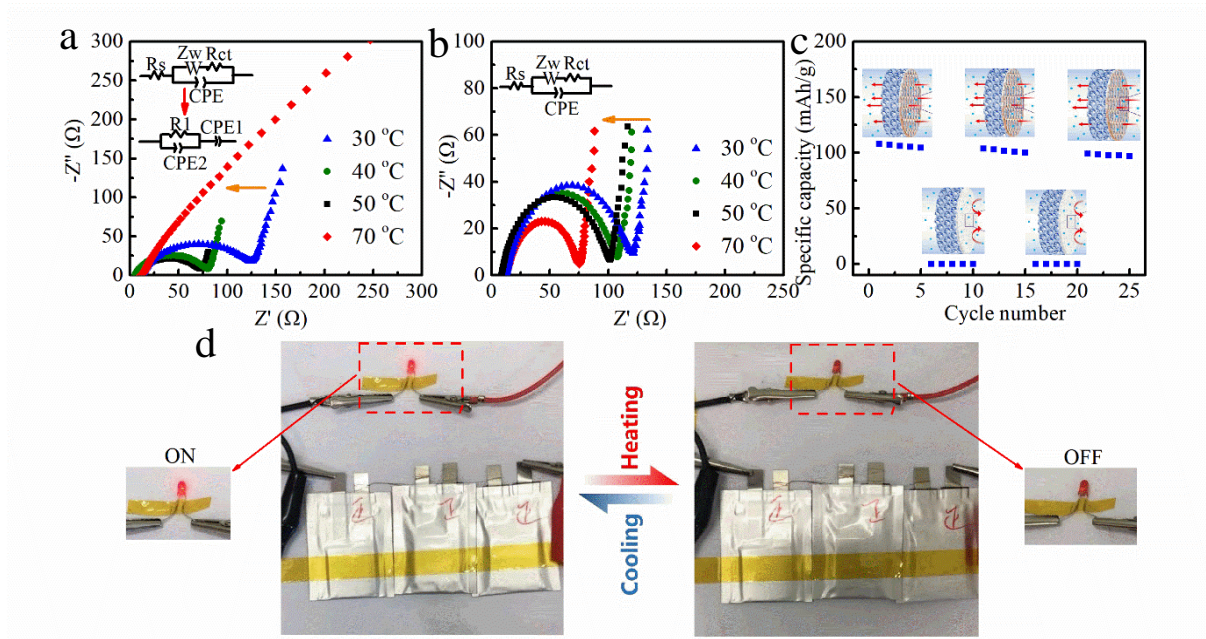

Fig. 6 (a) Temperature dependent EIS of the full battery with a glass-fiber separator. (b) Temperature dependent EIS of the full battery with a thermoresponsive separator. (c) Self-protection in several heating-cooling cycles. (d) Picture of conversion of led lighting and shut-off.

To directly illustrate the self-protection ability of the hydrogel-modified separator, the specific capacity of the battery was recorded for several consecutive heating and cooling cycles, as shown in (Fig. 6c). At high temperatures, the specific capacity of the aqueous lithium-ion battery decreased to 0 and realized the shut-off. When the battery cooled down, the specific capacity could recover to the normal capacity, ~100 mAh/g. To further demonstrate the self-protection behaviour in a battery pack. Three pouch-cell batteries were connected in series to power a light-emitting diode (LED) (Fig. 6d). The LED bulb was on at the beginning; however, the LED was turned off when the temperature rises to 70 $^{o}$C, indicating the self-protection ability of the modified separator. Subsequently cooling the battery to room temperature, the LED was lit up again and as bright as before. These results successfully verify the feasibility of self-protection ability and good thermoresponsive reversibility of the modified separator in aqueous lithium-ion batteries.

In conclusion, we developed smart inherently self-protecting aqueous lithium-ion batteries using thermoresponsive separators, which is synthesized by in-situ polymerization of hydrogels on the surface of hydrophilic fiber. The growth of hydrogels on the porous and hydrophilic matrix makes it possible to tightly coat a thin and uniform layer on the hydrophilic separator due to the low surface tension. The closed channel and hydrophobicity of the modified separator cut off the transport of ions between cathodes and anodes at high temperature to achieve self-protection, while the battery returns to the normal state due to reopened channel and hydrophilicity of the separator when it cools down. Impressively, the whole process is completely reversible. In addition, the thermoresponsive property is seriously deteriorated in kosmotropes salts solution but keeps almost the same in chaotropes salts solution. Meanwhile, the chaotropes salt, $LiNO_3$, is one of the best candidates for aqueous lithium-ion batteries. Therefore, for the first time, reversible thermoresponsive self-protection was achieved without sacrificing the battery performance in aqueous lithium-ion batteries using $LiNO_3$. In conclusion, this work provides effective and smart approaches to extend lifespan and ensure safety of aqueous batteries.